\documentclass[
]{ceurart}

\usepackage{listings}
\usepackage{cleveref}
\usepackage{hyperref} 
\usepackage[acronym]{glossaries}
\usepackage{tikz}
\usetikzlibrary{positioning, arrows.meta, shapes.geometric}
\tikzset{
    every node/.style={font=\small},
    io/.style={trapezium, trapezium left angle=70, trapezium right angle=110, draw, minimum height=1cm},
    block/.style={rectangle, draw, minimum height=1cm, minimum width=2.2cm, align=center},
    qc/.style={rectangle, draw, fill=blue!10, minimum height=1cm, minimum width=2.2cm, align=center},
    qout/.style={rectangle, draw, dashed, minimum height=1cm, minimum width=2.2cm, align=center},
    arrow/.style={draw, thick, -{Latex[length=3mm]}},
}

\crefname{table}{Tab.}{Tabs.}
\newacronym{qce}{QCE}{Quantum Circuit Expressibility}
\newacronym{qcf}{QCF}{Quantum Circuit Fidelity}

\newacronym{qlr}{QLR}{Quantum Locality Ratio}
\newacronym{eee}{EEE}{Effective Entanglement Entropy}
\newacronym{qmi}{QMI}{Quantum Mutual Information}

\newacronym{fmcr}{FMCR}{Feature Map Compression Ratio}
\newacronym{edqfs}{EDQFS}{Effective Dimension of Quantum Feature Space}
\newacronym{qlad}{QLAD}{Quantum Layer Activation Diversity}
\newacronym{qos}{QOS}{Quantum Output Sensitivity}
\newacronym{tsi}{TSI}{Training Stability Index}
\newacronym{tei}{TEI}{Training Efficiency Index}
\newacronym{qgn}{QGN}{Quantum Gradient Norm}
\newacronym{bpi}{BPI}{Barren Plateau Indicator}
\newacronym{rqlsi}{RQLSI}{Relative Quantum Layer Stability Index}
\newacronym{rqtei}{r-QTEI}{Relative Quantum Training Efficiency Index}

\begin{document}

\copyrightyear{2025}
\copyrightclause{Copyright for this paper by its authors.
  Use permitted under Creative Commons License Attribution 4.0
  International (CC BY 4.0).}

\conference{The Fourth Conference on System and Service Quality - QualITA, June 25-26, 2025, Catania, Italy}

\title{QMetric: Benchmarking Quantum Neural Networks Across Circuits, Features, and Training Dimensions}

\author[1]{Silvie Illésová}[%
orcid=0009-0002-5231-3714,
email=illesova.silvie.scholar@gmail.com,
]
\cormark[1]
\fnmark[1]

\address[1]{IT4Innovations National Supercomputing Center,
Studentská 6231/1B,
708 00 Ostrava,
Czech Republic}

\author[2,3,4]{Tomasz Rybotycki}[%
orcid=0000-0003-2493-0459,
email=tryb@camk.edu.pl,
]
\fnmark[1]

\address[2]{Systems Research Institute, Polish Academy of Sciences, ul. Newelska 6, 01-447 Warszawa, Poland}
\address[3]{ Nicolaus Copernicus Astronomical Center, Polish Academy of Sciences, ul. Bartycka 18, 00-716 Warsaw, Poland}
\address[4]{Center of Excellence in Artificial Intelligence, AGH University, Aleje Mickiewicza 30, 30-059 Cracow, Poland}

\author[5]{Martin Beseda}[%
orcid=0000-0001-5792-2872,
email=martin.beseda@univaq.it,
]
\fnmark[1]

\address[5]{Department of Information Engineering, Computer Science and Mathematics, University of L'Aquila, via Vetoio,
I-67010 Coppito-L'Aquila, Italy}

\newcommand{\mb}[1]{{\color{blue}MB: #1}}

\cortext[1]{Corresponding author.}
\fntext[1]{These authors contributed equally.}

\begin{abstract}
As hybrid quantum-classical models gain traction in machine learning, there is a growing need for tools that assess their effectiveness beyond raw accuracy. We present \textbf{QMetric}, a Python package offering a suite of interpretable metrics to evaluate quantum circuit expressibility, feature representations, and training dynamics. QMetric quantifies key aspects such as circuit fidelity, entanglement entropy, barren plateau risk, and training stability. The package integrates with Qiskit and PyTorch, and is demonstrated via a case study on binary MNIST classification comparing classical and quantum-enhanced models. Code, plots, and a reproducible environment are available on GitLab.
\end{abstract}



\maketitle

\section{Introduction}
Hybrid quantum-classical neural networks (QNNs) \cite{liu2021hybrid,chen2024deep,zeng2022multi} are playing a central role in the development of algorithms for near-term quantum devices. By embedding parameterized quantum circuits within classical training loops, these architectures aim to leverage quantum resources such as entanglement and superposition while maintaining trainability through well-established classical optimizers. This hybrid structure has enabled a broad spectrum of quantum machine learning (QML) \cite{biamonte2017quantum,schuld2015introduction} models to flourish across domains including classification \cite{senokosov2024quantum,lu2024quantum}, generative modeling\cite{riofrio2024characterization,coyle2021quantum}, optimization \cite{divya2021quantum,al2025optimization,novak2025optimization}, benchmarking\cite{bilek2025experimental,lewandowska2025benchmarking}, medicine\cite{novak2025predicting,wei2023quantum,rani2023quantum,thiyagalingam2022scientific}, and quantum chemistry\cite{sajjan2022quantum,huang2020quantum,bauer2025efficient}.

Importantly, many canonical variational algorithms—originally developed for quantum simulation—can be reframed as learning architectures. The \emph{Variational Quantum Eigensolver} (VQE) \cite{fedorov2022vqe, illesova2025numerical,ciaramelletti2025,zhang2025qracle} exemplifies this duality a parameterized quantum ansatz is trained to minimize an energy objective, just like a neural network minimizing a loss function. Over time, VQE has evolved into a family of learning-based formulations, including \emph{State-Averaged Orbital-Optimized VQE} \cite{beseda2024state, illesova2025transformation}, \emph{ADAPT-VQE} \cite{tang2021qubit}, and \emph{Subspace-Search VQE} \cite{nakanishi2019subspace}, each introducing novel strategies for parameterization, target state selection, and optimization flow.

Beyond simulation, hybrid QNNs are widely applied in supervised learning, typically in classification or regression tasks\cite{gupta2022quantum}. Here, quantum circuits are used to encode classical data (via feature maps \cite{suzuki2020analysis,kwon2024feature}), process it through a variational ansatz, and output probabilities or decision boundaries. Such architectures are used in \emph{Quantum Neural Networks} \cite{gupta2001quantum}, \emph{Quantum Support Vector Machines} \cite{rebentrost2014quantum}, and more recent paradigms like \emph{Quantum Kitchen Sinks} \cite{wilson2018quantum} and \emph{Quantum Feature Spaces} \cite{havlivcek2019supervised}. In unsupervised learning, models such as \emph{Quantum Circuit Born Machines} \cite{liu2018differentiable}, and \emph{Quantum Autoencoders} \cite{locher2023quantum} extend the reach of QML into generative and latent-variable modeling.

Despite the growing variety and sophistication of QML models, there remains a lack of principled, interpretable, and reproducible tools for evaluating their behavior. Traditional ML diagnostics—accuracy, F1-score, or validation loss—do not capture key quantum characteristics such as circuit expressibility, entanglement structure, barren plateaus, or the sensitivity of quantum feature maps. Without such metrics, model design becomes largely heuristic and comparisons between quantum and classical architectures are often inconclusive or misleading.

To bridge this gap, we introduce \textbf{QMetric}, a modular and extensible Python framework for evaluating hybrid quantum-classical models. QMetric computes interpretable scalar metrics across three complementary dimensions (i) the structure and expressiveness of quantum circuits; (ii) the geometry and compression of quantum feature spaces; and (iii) the stability, efficiency, and gradient flow during training. These tools allow researchers to diagnose bottlenecks, compare architectures, and validate empirical claims beyond raw accuracy.

Our package integrates with Qiskit \footnote{https://www.ibm.com/quantum/qiskit} and PyTorch \footnote{https://pytorch.org/} which demonstrated through a binary classification example on MNIST digits, comparing a classical neural network to a hybrid QNN. All code, plots, and environment files are publicly available for reproducibility and further experimentation.

\section{Software Specifications}

All experiments were conducted using the \texttt{qmetric-env} Conda \footnote{https://anaconda.org/anaconda/conda} environment \footnote{https://gitlab.com/illesova.silvie.scholar/qmetric/-/blob/main/environment.yml}, configured for hybrid quantum-classical machine learning. The system exploits GPU-accelerated libraries, supports Qiskit primitives of version V1, and integrates PyTorch and scikit-learn\footnote{https://scikit-learn.org/stable/} for classical model components and preprocessing.

The environment is based on Python 3.10.13 with key libraries and versions listed in Table~\ref{tab:software}. Qiskit version 1.4.3 was used in conjunction with Qiskit Aer 0.17.0 and Qiskit Machine Learning 0.8.2. PyTorch version 2.7.0 and CUDA 12.9 toolchain were used for classical and hybrid model execution. Principal component analysis and classical baseline training relied on scikit-learn version 1.6.1.

\begin{table}[h]
\centering
\caption{Key Software Components}
\label{tab:software}
\begin{tabular}{ll}
\toprule
\textbf{Library} & \textbf{Version} \\
\midrule
Python & 3.10.13 \\
Qiskit & 1.4.3 \\
Qiskit Aer & 0.17.0 \\
Qiskit Machine Learning & 0.8.2 \\
PyTorch & 2.7.0 \\
CUDA Toolkit & 12.9 \\
cuDNN & 9.10.1.4 \\
scikit-learn & 1.6.1 \\
NumPy & 2.2.6 \\
Matplotlib & 3.10.3 \\
SymPy & 1.14.0 \\
\bottomrule
\end{tabular}
\end{table}

The training experiments were run on a Linux system using conda’s \texttt{qmetric-env} environment. Hardware acceleration via CUDA and cuDNN was enabled to support efficient execution of neural network operations and gradient computation. Quantum simulations were executed using \texttt{AerSimulator} in statevector mode.

Note that the Qiskit primitives interface used in the hybrid model, \texttt{EstimatorQNN}\footnote{https://qiskit-community.github.io/qiskit-machine-learning/stubs/qiskit\_machine\_learning.neural\_networks.EstimatorQNN.html}, is marked as deprecated in favor of V2 primitives. Future implementations of QMetric should migrate to the \texttt{StatevectorEstimator}\footnote{https://quantum.cloud.ibm.com/docs/en/api/qiskit/qiskit.primitives.SamplerPubResult} to ensure compatibility with upcoming Qiskit releases.
\section{Metrics Categories}

QMetric organizes its metrics into three complementary categories—quantum circuit behavior, quantum feature space, and training dynamics—that together provide a comprehensive profile of a hybrid model’s expressiveness, learnability, and robustness. These categories are summarised in \cref{tab:qmetric-condensed}.

\subsection{Quantum Circuit Metrics}

As the computational core of hybrid models, quantum circuits influence representational capacity and noise resilience. QMetric evaluates circuit quality through metrics such as \emph{Quantum Circuit Expressibility}, which measures the diversity of quantum states produced under random parameters, and \emph{Quantum Circuit Fidelity}, estimating robustness to noise via state overlap.

To characterize circuit structure, the \emph{Quantum Locality Ratio} captures the balance between local and entangling gates. Entanglement-based metrics include \emph{Effective Entanglement Entropy} and \emph{Quantum Mutual Information}, which quantify intra-circuit quantum correlations. These metrics are useful when tuning ansätze for VQE, QAOA, or classification tasks, where poor expressibility or excessive entanglement can hinder learning.

\subsubsection{Quantum Circuit Expressibility}
\gls{qce} \cite{zhang2024transformer} quantifies a circuit's ability to generate a diverse set of quantum states across the Hilbert space. It measures how closely the distribution of states produced by the parameterized circuit approximates the uniform (Haar) distribution \cite{ter1955foundations}. High expressibility corresponds to broader state coverage in Hilbert space and is conceptually linked to the Fubini-Study distance \cite{yang2021gravity} and can be quantitatively related to the Kullback–Leibler divergence \cite{kullback1951kullback} between the circuit's output distribution and the Haar distribution. It implies that the circuit can reach a wide variety of states, which is crucial for representing complex functions in quantum machine learning and variational algorithms.

Formally, \gls{qce} is defined via the pairwise fidelity of randomly generated statevectors,
\begin{equation}
  \text{QCE} = 1 - \frac{1}{N(N-1)} \sum_{i < j} \left| \langle \psi_i | \psi_j \rangle \right|^2,  
\end{equation}
where $N$ is the number of randomly sampled parameter sets used to generate the corresponding quantum states, $\{ \psi_i \}_{i=1}^N$ are the quantum states obtained by randomly sampling parameters from the specified ranges and applying them to the circuit. This expression captures the average overlap between states, lower overlap corresponds to greater expressibility. The \gls{qce} score lies in the range $[0, 1]$, with values closer to 1 indicating higher expressiveness.

In practice, \gls{qce} helps identify whether a variational circuit is too shallow (low expressibility) or overly complex (potentially prone to barren plateaus). A well-designed circuit should maintain a high \gls{qce} while preserving trainability and manageable entanglement. QMetric implements \gls{qce} by sampling multiple parameter sets, evaluating statevector overlaps, and averaging across all pairwise fidelities, making it an efficient diagnostic for early-stage ansatz evaluation.

\subsubsection{Quantum Circuit Fidelity}

\gls{qcf}  \cite{mendoncca2008alternative} quantifies the robustness of a quantum circuit to noise by measuring how closely the output of the noisy circuit resembles that of the ideal (noise-free) version. Fidelity serves as a key metric for assessing noise resilience in near-term quantum devices, where decoherence, gate errors, and readout noise can significantly degrade quantum state quality.

Mathematically, \gls{qcf} is defined as the fidelity between two quantum states,
\begin{equation}
    F(\rho, \sigma) = \left( \text{Tr} \sqrt{\sqrt{\rho} \sigma \sqrt{\rho}} \right)^2,
\end{equation}

where $\rho$ is the density matrix of the ideal output state and $\sigma$ represents the output state under a specified noise model. In the special case of pure states (as in most simulation scenarios), the fidelity simplifies to the squared absolute value of the inner product between the ideal and noisy statevectors.

In QMetric, \gls{qcf} is computed by simulating both the ideal and noisy execution of a circuit using Qiskit’s statevector simulator and a user-defined noise model. The resulting fidelity score ranges from 0 to 1, with higher values indicating stronger fidelity. \gls{qcf} is especially useful when benchmarking circuits across different hardware targets or when optimizing ansatz designs for noisy intermediate-scale quantum (NISQ) devices.

\subsubsection{Quantum Locality Ratio}

\gls{qlr} \cite{bertini2020operator} quantifies the proportion of single-qubit operations relative to the total number of gates in a quantum circuit. This metric captures the locality of interactions, offering insight into how much a circuit relies on entangling operations. A high \gls{qlr} implies that the circuit uses mostly local, single-qubit gates, whereas a low value suggests strong reliance on multi-qubit entanglement.

Formally, \gls{qlr} is defined as
\begin{equation}
\text{QLR} = \frac{N_{\text{1-q}}}{N_{\text{total}}},
\end{equation}
where $ N_{\text{1-q}}$ denotes the number of gates acting on a single qubit and $ N_{\text{total}} $ is the total number of gates in the circuit.

In QMetric, this ratio is computed by iterating over the circuit’s gate operations and counting how many act on one qubit. \gls{qlr} helps to assess the tradeoff between locality and entanglement, basically it tells us the ration of single to multi-qubit gates, providing a fast and interpretable structural descriptor. It is particularly useful during ansatz design where excessive entanglement can introduce barren plateaus or hardware noise sensitivity.

\subsubsection{Effective Entanglement Entropy}

\gls{eee} \cite{boes2019neumann} evaluates the degree of quantum entanglement between a subsystem of qubits and the rest of the circuit. It is based on the von Neumann entropy of the reduced density matrix of the selected subsystem, capturing how mixed its state becomes due to entanglement with its complement.

The metric is defined as
\begin{equation}
S(\rho_A) = -\text{Tr}(\rho_A \log \rho_A),
\end{equation}
where $ \rho_A $ is the reduced density matrix obtained by tracing out all qubits not in the chosen subsystem.

QMetric computes \gls{eee} by generating a statevector from the circuit, selecting a target subset of qubits, performing partial trace, and evaluating the entropy. This metric is useful in tasks like entanglement scaling analysis where understanding subsystem correlations is essential for tuning circuit depth and topology.

\subsubsection{Quantum Mutual Information}

\gls{qmi} \cite{schumacher2006quantum} measures the total correlations—both classical and quantum—between two disjoint subsets of qubits in a quantum circuit. It extends the concept of mutual information to the quantum domain, revealing how strongly two regions of a circuit are statistically linked.

The metric is computed as
\begin{equation}
I(A:B) = S(\rho_A) + S(\rho_B) - S(\rho_{AB}),
\end{equation}
where $ \rho_A $, $ \rho_B $, and $ \rho_{AB} $ are the reduced density matrices of subsystems $ A $, $ B $, and their union, respectively.

In QMetric, \gls{qmi} is calculated by preparing a full statevector, currently via analytical simulation, computing partial traces for each subsystem and their union, and evaluating the entropies. This metric is instrumental for analyzing modular architectures, verifying disentanglement, or diagnosing undesired correlations in VQE, QAOA, or classification-oriented quantum circuits.

\subsection{Quantum Feature Space Metrics}

When encoding classical data into Hilbert space, the geometry of the resulting feature space directly affects model performance. QMetric provides the \emph{Feature Map Compression Ratio} (FMCR), assessing how efficiently classical data are compressed via PCA, and the \emph{Effective Dimension} (EDQFS), which reflects variance spread in the quantum feature space.

The \emph{Quantum Layer Activation Diversity} (QLAD) and \emph{Quantum Output Sensitivity} (QOS) evaluate output variability and robustness to perturbations. Low QLAD and high QOS signal collapsed or brittle encodings. These metrics are critical in PQC-based classifiers, quantum kernel methods, and other models relying on quantum feature geometry.

\subsubsection{Feature Map Compression Ratio}

\gls{fmcr} \cite{mackiewicz1993principal} quantifies how efficiently a quantum feature map compresses the input data. It compares the original classical dimensionality with the number of principal components needed to capture most of the variance in the quantum-transformed space. A high \gls{fmcr} indicates strong compression, meaning fewer effective dimensions are required to retain the majority of the encoded information.

Formally, it is defined as
\begin{equation}
\text{FMCR} = \frac{d_{\text{in}}}{d_{\text{eff}}},
\end{equation}
where $ d_{\text{in}} $ is the dimensionality of the classical input and $ d_{\text{eff}} $ is the number of principal components explaining 95\% of the variance in the quantum feature space.

QMetric implements \gls{fmcr} by applying PCA to the quantum-transformed dataset, calculating the cumulative explained variance, and identifying the number of components required to exceed the 95\% threshold. This metric is especially relevant when assessing whether a feature map leads to redundancy or useful abstraction.

\subsubsection{Effective Dimension of Quantum Feature Space}

\gls{edqfs} \cite{crossley1976effective} measures how uniformly information is distributed in the quantum feature space. It is based on the PCA eigenvalue spectrum and captures the intrinsic dimensionality of the embedded data. A high \gls{edqfs} suggests a flat eigenvalue distribution and a more balanced use of the available Hilbert space dimensions.

The effective dimension is computed as
\begin{equation}
d_{\text{eff}} = \frac{\left(\sum_i \lambda_i\right)^2}{\sum_i \lambda_i^2},
\end{equation}
where $ \lambda_i $ are the PCA eigenvalues of the quantum-encoded dataset, that was encoded by feature map. The summation runs over all principal components, i.e., $i = 1, \dots, r$, where $r = \min(n, d)$ is the rank of the dataset with $n$ samples and $d$ features.

QMetric calculates \gls{edqfs} by performing PCA on the quantum features and evaluating the above formula. This metric complements \gls{fmcr} by indicating how efficiently the encoded dimensions are utilized, helping to diagnose over- or under-spread feature distributions.

\subsubsection{Quantum Layer Activation Diversity}

\gls{qlad} \cite{xie2017diverse} evaluates the diversity of measurement outcomes across samples in the quantum feature space. It is based on the variance of probability distributions obtained from quantum measurements, reflecting how varied the output activations are for different inputs.

The metric is defined as
\begin{equation}
\text{QLAD} = \frac{1}{n} \sum_{i=1}^n \text{Var}(p_i),
\end{equation}
where $ p_i $ is the measurement probability distribution for the $ i $-th sample and $ n $ is the number of samples.

In QMetric, \gls{qlad} is computed by estimating the variance across each sample’s probability vector and averaging the results. Low \gls{qlad} may signal that the circuit is collapsing inputs into narrow output distributions, which could hinder expressivity and classification power.

\subsubsection{Quantum Output Sensitivity}

\gls{qos} \cite{liu2024jacobian} measures how sensitive a quantum model’s output is to small perturbations in the input. It captures robustness and smoothness of the mapping from classical data to quantum measurements. A low \gls{qos} implies a stable, noise-tolerant model, while a high value may indicate fragility or sharp decision boundaries.

The metric is computed as
\begin{equation}
\text{QOS} = \mathbb{E} \left[ \frac{\|f(x + \epsilon) - f(x)\|^2}{\|\epsilon\|^2} \right],
\label{eq:E}
\end{equation}
where $ f(x) $ and $ f(x + \epsilon) $ are the quantum model outputs for the original and perturbed inputs, respectively. Here, $\mathbb{E}$ denotes the empirical average over a batch of perturbation vectors $\epsilon$, typically sampled from a zero-mean isotropic Gaussian distribution.

In QMetric, \gls{qos} is evaluated by generating perturbed versions of inputs, computing the model output differences, and normalizing by the squared perturbation norms. This metric is useful for analyzing encoding smoothness, adversarial stability, and overall model resilience.

\begin{table}[h!]
\centering
\caption{Summary of QMetric Metrics}
\begin{tabular}{|l|l|p{6.5cm}|}
\hline
\textbf{Category} & \textbf{Metric} & \textbf{Purpose} \\
\hline
\multirow{3}{*}{Quantum Circuit} 
  & QCE, QCF & Expressibility and noise robustness of circuits. \\
  & QLR & Balance of local vs. entangling gates. \\
  & EEE, QMI & Entanglement and intra-circuit correlation. \\
\hline
\multirow{3}{*}{Feature Space} 
  & FMCR, EDQFS & Data compression and feature spread \\
  & QLAD & Diversity of quantum activations. \\
  & QOS & Sensitivity to small input changes. \\
\hline
\multirow{3}{*}{Training Dynamics} 
  & TSI, TEI & Stability and efficiency of convergence. \\
  & QGN, BPI & Gradient health and barren plateau risk. \\
  & RQLSI, r-QTEI & Relative diagnostics vs classical models. \\
\hline
\end{tabular}
\label{tab:qmetric-condensed}
\end{table}

\subsection{Training Dynamics}

QMetric also tracks training behavior using the \emph{Training Stability Index} (TSI), which compares variability in training and validation losses, and the \emph{Training Efficiency Index} (TEI), which measures epochs needed to reach a target accuracy relative to model size.

Quantum-specific diagnostics include the \emph{Quantum Gradient Norm} (QGN) and \emph{Barren Plateau Indicator} (BPI), both of which expose vanishing gradients linked to deep or poorly initialized circuits. To compare hybrid and classical models, relative metrics such as \emph{RQLSI} and \emph{r-QTEI} quantify differences in training efficiency and stability under aligned conditions. Together, these metrics support targeted diagnosis of underperformance, guide ansatz design, and enable meaningful evaluation across model types.

\subsubsection{Training Stability Index}

\gls{tsi} \cite{chandramoorthy2022generalization}quantifies the variability in training and validation losses near convergence. It measures how consistently the model performs in the final training phase by comparing the standard deviation of losses over the last 10\% of epochs, this percentage will be up to the user in the future versions of QMetric.

The metric is defined as
\begin{equation}
\text{TSI} = \frac{\sigma_{\text{val}}}{\sigma_{\text{train}}},
\end{equation}
where $ \sigma_{\text{train}} $ and $ \sigma_{\text{val}} $ denote the standard deviations of training and validation losses, respectively. Here, "losses" refer to the recorded values of the loss function (e.g., cross-entropy or MSE) over training and validation batches during training.

A low \gls{tsi} indicates stable and consistent generalization, while a high value may reveal overfitting or noisy training dynamics. In QMetric, \gls{tsi} is computed by evaluating standard deviations over the tail segment of the loss curves, i.e. we take into account the standart deviation of the last $n$ outputs of the loss function.

\subsubsection{Training Efficiency Index}

\gls{tei} \cite{livni2014computational} measures how quickly a model reaches a high level of performance relative to its size. It is defined as the ratio between the epoch at which validation accuracy first exceeds 90\% and the number of trainable parameters.

Formally,
\begin{equation}
\text{TEI} = \frac{\text{epoch}_{90}}{N_{\text{params}}},
\end{equation}
where $ \text{epoch}_{90} $, again the threshold will be up to the user in future version, is the earliest epoch where 90\% accuracy is reached and $ N_{\text{params}} $ is the total parameter count.

Lower \gls{tei} values indicate faster convergence per parameter, making this metric a useful tool for evaluating training efficiency across differently sized architectures.

\subsubsection{Quantum Gradient Norm}

\gls{qgn} \cite{galegale2024deep} measures the magnitude of gradients associated with quantum circuit parameters. It reflects the overall strength of parameter updates and can signal the presence of vanishing or exploding gradients.

The metric is computed as
\begin{equation}
\text{QGN} = \left\| \nabla_{\theta_q} \mathcal{L} \right\|_2,
\end{equation}
where $ \theta_q $ denotes the quantum parameters and $ \mathcal{L} $ is the training loss, i.e. the value of loss function on the training dataset.

In QMetric, gradients are extracted from the backpropagation step and concatenated for L2 norm calculation
\begin{equation}
\|\nabla \mathcal{L}\|_2 = \left\| \left[ \frac{\partial \mathcal{L}}{\partial \theta_1}, \frac{\partial \mathcal{L}}{\partial \theta_2}, \dots, \frac{\partial \mathcal{L}}{\partial \theta_n} \right] \right\|_2,
\end{equation}
where $\mathcal{L}$ is the loss function and $\{ \theta_i \}_{i=1}^n$ are the trainable parameters of the hybrid quantum-classical model. Low \gls{qgn} may indicate a barren plateau or excessively deep circuits.

\subsubsection{Barren Plateau Indicator}

\gls{bpi} \cite{larocca2022diagnosing} estimates whether a model suffers from barren plateaus by evaluating the average squared magnitude of quantum gradients. This captures the extent of vanishing gradients during optimization.

It is defined as
\begin{equation}
\text{BPI} = \mathbb{E} \left[ \left\| \nabla_{\theta_q} \mathcal{L} \right\|^2 \right],
\end{equation}

where $\mathbb{E}$ is the same as in \cref{eq:E}. Values near zero suggest that gradients are vanishing, which can hinder effective training. In QMetric, \gls{bpi} is computed over the flattened list of quantum gradients and averaged into final value, making it an efficient early diagnostic tool during model tuning.

\subsubsection{Relative Quantum Layer Stability Index}

\gls{rqlsi} \cite{liuunderstanding} compares the training stability of hybrid quantum-classical models to that of purely classical ones using the \gls{tsi} metric. It helps quantify whether introducing quantum layers improves or worsens loss stability.

Formally,
\begin{equation}
\text{RQLSI} = \frac{\text{TSI}_{\text{hybrid}}}{\text{TSI}_{\text{classical}}},
\end{equation}
where $ \text{TSI}_{\text{hybrid}} $ and $ \text{TSI}_{\text{classical}} $ are the training stability indices for the hybrid and classical models, respectively.

A value less than 1 suggests that the quantum-enhanced model is more stable during training. This metric supports empirical comparison between model types under matched conditions.

\subsubsection{Relative Quantum Training Efficiency Index}

\gls{rqtei} \cite{tan2021efficientnetv2} evaluates whether a hybrid model trains more efficiently than a classical counterpart by comparing their respective \gls{tei} scores.

It is defined as
\begin{equation}
\text{r-QTEI} = \frac{\text{TEI}_{\text{hybrid}}}{\text{TEI}_{\text{classical}}},
\end{equation}
where $ \text{TEI}_{\text{hybrid}} $ and $ \text{TEI}_{\text{classical}} $ are the training efficiency indices.

A value below 1 means the hybrid model reaches target performance faster relative to its parameter size. In QMetric, this metric supports head-to-head benchmarking of model variants in practical scenarios.

\section{Case Study: Hybrid vs Classical on MNIST}

To illustrate the diagnostic capabilities of QMetric, we evaluate a hybrid quantum-classical neural network against a classical baseline using a binary classification task on the MNIST dataset \cite{deng2012mnist}. This case study provides a practical scenario where quantum neural networks are tested under realistic constraints. We describe the model architectures, data pipeline, training configuration, and metric-driven analysis.

\subsection{Hybrid Model}

The hybrid model integrates a parameterized quantum circuit with a classical output layer to perform binary classification. The quantum component is implemented using Qiskit’s \texttt{EstimatorQNN} and is connected to PyTorch via the \texttt{TorchConnector}, allowing seamless integration with PyTorch’s autograd system.

The quantum circuit is composed of a feature map and an ansatz. The feature map is a \texttt{ZZFeatureMap} with one repetition that encodes classical inputs into quantum states. The ansatz is a \texttt{RealAmplitudes} circuit with three repetitions that introduces trainable parameters and entanglement. These circuits are composed into a single parameterized circuit which is then used to define the quantum neural network. The circuit outputs a single expectation value which is passed through a trainable classical linear layer followed by a sigmoid activation. The full model maps input vector $x$ to output $\sigma(W \cdot \text{QNN}(x) + b)$ where $W$ and $b$ are trainable classical parameters.

To match the number of qubits in the circuit, the MNIST images are projected into a lower-dimensional space using principal component analysis. The original 784-dimensional vectors are reduced to three components. This projection ensures compatibility with a three-qubit quantum circuit while preserving as much variance as possible.

The dataset is constructed by filtering the MNIST training set to include only samples corresponding to digits 0 and 1. From this filtered subset, the first 500 examples are selected to simulate a small-data regime. The images are flattened into vectors, normalized, and then transformed using PCA to produce a dataset suitable for quantum encoding.

The hybrid model is trained for 30 epochs using the Adam optimizer with a learning rate of 0.01. Binary cross-entropy is used as the loss function. Training and validation losses are tracked at each epoch along with validation accuracy. Additionally, gradients with respect to the quantum parameters are collected to enable computation of metrics such as the quantum gradient norm and barren plateau indicator.

After training, the quantum outputs are evaluated using QMetric. Metrics such as the feature map compression ratio, effective dimension of the quantum feature space, layer activation diversity, and output sensitivity are computed from the post-quantum activations. Quantum circuit diagnostics such as expressibility, locality ratio, entanglement entropy, mutual information, and noise robustness are also evaluated. This model provides a complete use case for applying QMetric during model selection, architectural tuning, and training analysis.

\begin{figure}[h]
\centering
\begin{tikzpicture}[node distance=0.4cm]

\node[block] (input) {Classical Input \\ (3D vector)};
\node[qc, below=of input] (fm) {Quantum Feature Map \\ (ZZFeatureMap)};
\node[qc, below=of fm] (ansatz) {Parameterized Circuit \\ (RealAmplitudes)};
\node[qout, below=of ansatz] (qout) {Expectation Value};
\node[block, below=of qout] (fc) {Fully Connected Layer};
\node[block, below=of fc] (sigmoid) {Sigmoid Activation};
\node[block, below=of sigmoid] (output) {Prediction \\ (0 or 1)};

\draw[arrow] (input) -- (fm);
\draw[arrow] (fm) -- (ansatz);
\draw[arrow] (ansatz) -- (qout);
\draw[arrow] (qout) -- (fc);
\draw[arrow] (fc) -- (sigmoid);
\draw[arrow] (sigmoid) -- (output);

\end{tikzpicture}
\caption{Hybrid quantum-classical model architecture. Classical inputs are encoded into quantum states via a feature map and processed by a parameterized circuit. The quantum output is passed to a classical linear layer and sigmoid activation for binary classification.}
\end{figure}
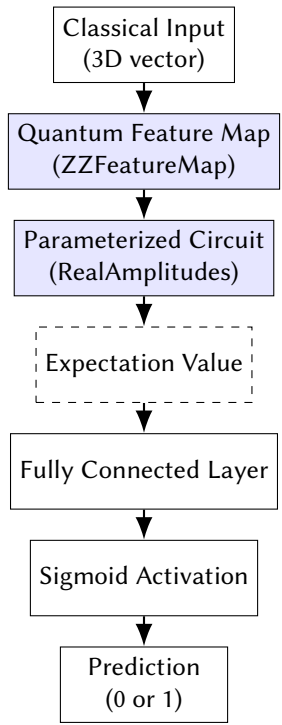

\subsection{Classical Baseline}

The classical baseline model is a fully connected neural network designed to match the input dimensionality and output behavior of the hybrid model. It takes as input the same three-dimensional data produced by PCA and outputs a binary classification probability using a sigmoid activation.

The architecture consists of an input layer with three neurons, a hidden layer with ten neurons using the ReLU activation function, and an output layer with one neuron followed by a sigmoid activation. The network approximates a function $f(x) = \sigma(W_2 \cdot \text{ReLU}(W_1 x + b_1) + b_2)$ where $x$ is the PCA-reduced input vector and $W_1$, $W_2$, $b_1$, and $b_2$ are trainable parameters.

The model is trained using the same subset of the MNIST dataset as the hybrid model. The inputs are 500 grayscale images corresponding to digits 0 and 1, flattened and reduced to three principal components. The preprocessing pipeline is identical, ensuring a fair comparison in terms of input dimensionality and task complexity.

The training procedure mirrors that of the hybrid model. The optimizer is Adam with a learning rate of 0.01, the loss function is binary cross-entropy, and the number of training epochs is set to 30. At each epoch, training loss, validation loss, and validation accuracy are recorded to allow for direct comparison of convergence dynamics, generalization performance, and learning stability.

This classical model serves as a baseline for interpreting the added value or limitations of quantum components under identical data, dimensionality, and optimization conditions. It enables a controlled analysis of the effects of quantum layers on expressivity, robustness, and trainability using QMetric's evaluation framework.
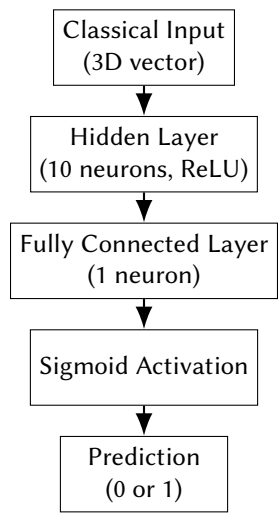
\begin{figure}[h]
\centering
\begin{tikzpicture}[node distance=0.4cm]

\node[block] (input) {Classical Input \\ (3D vector)};
\node[block, below=of input] (hidden) {Hidden Layer \\ (10 neurons, ReLU)};
\node[block, below=of hidden] (fc) {Fully Connected Layer \\ (1 neuron)};
\node[block, below=of fc] (sigmoid) {Sigmoid Activation};
\node[block, below=of sigmoid] (output) {Prediction \\ (0 or 1)};

\draw[arrow] (input) -- (hidden);
\draw[arrow] (hidden) -- (fc);
\draw[arrow] (fc) -- (sigmoid);
\draw[arrow] (sigmoid) -- (output);

\end{tikzpicture}
\caption{Architecture of the classical baseline neural network. A fully connected feedforward model processes PCA-reduced MNIST inputs to perform binary classification between digits 0 and 1.}
\end{figure}

\subsection{Quantum Circuit Metrics}

Table~\ref{tab:circuit-metrics} summarizes the metrics that characterize the structure, expressibility, and robustness of the quantum circuit used in the hybrid model.

\begin{table}[h]
\centering
\caption{Quantum Circuit Metrics}
\label{tab:circuit-metrics}
\begin{tabular}{lcc}
\toprule
\textbf{Metric} & \textbf{Value} & \textbf{Interpretation} \\
\midrule
Quantum Circuit Expressibility (QCE) & 0.939 & High expressibility, close to Haar-random \\
Quantum Volume Contribution (QVC) & 0.1905 & Moderate depth, penalizing expressive capacity \\
Quantum Circuit Fidelity (QCF) & 1.000 & Perfect fidelity under simulation noise \\
Quantum Locality Ratio (QLR) & 0.6364 & Balanced mix of local and entangling gates \\
Effective Entanglement Entropy (EEE) & 0.8345 & High entanglement across qubit partitions \\
Quantum Mutual Information (QMI) & 1.6691 & Strong total correlations between subsystems \\
\bottomrule
\end{tabular}
\end{table}

The quantum circuit demonstrates high expressibility ($\text{QCE} = 0.939$), suggesting that it explores a diverse set of quantum states across the Hilbert space. The fidelity score ($\text{QCF} = 1.000$) confirms robustness to noise under simulation with a basic noise model. The locality ratio of $0.6364$ indicates a well-balanced design between local and entangling operations. Entanglement is both substantial and well-structured, as shown by high values of $\text{EEE} = 0.8345$ and $\text{QMI} = 1.6691$, supporting rich correlations necessary for quantum information processing.

\subsection{Feature Space Metrics}

The geometry and structure of the quantum feature space are assessed through the metrics in Table~\ref{tab:feature-metrics}, which evaluate compression, variance distribution, activation diversity, and sensitivity to perturbations.

\begin{table}[h]
\centering
\caption{Quantum Feature Space Metrics}
\label{tab:feature-metrics}
\begin{tabular}{lcc}
\toprule
\textbf{Metric} & \textbf{Value} & \textbf{Interpretation} \\
\midrule
Feature Map Compression Ratio (FMCR) & 3.000 & Strong compression from 3D input to 1D effective space \\
Effective Dimension (EDQFS) & 1.000 & Variance concentrated in a single direction \\
Quantum Layer Activation Diversity (QLAD) & 0.000 & Collapsed outputs, no diversity in activation patterns \\
Quantum Output Sensitivity (QOS) & 9.644 & Highly sensitive to small input perturbations \\
\bottomrule
\end{tabular}
\end{table}

The feature map achieves perfect compression ($\text{FMCR} = 3.0$), indicating that all input variance is concentrated in one effective principal component. However, the effective dimension ($\text{EDQFS} = 1.0$) confirms that the quantum feature space lacks spread. Activation diversity is entirely absent ($\text{QLAD} = 0.000$), signaling possible circuit over-regularization or symmetry that collapses measurement outputs. Meanwhile, the high sensitivity ($\text{QOS} = 9.644$) indicates the model reacts sharply to small perturbations, suggesting brittle or sharp decision boundaries.

\subsection{Training Dynamics}

Training dynamics of the hybrid and classical models are evaluated in Table~\ref{tab:training-metrics}. These metrics reflect convergence behavior, parameter efficiency, gradient stability, and vanishing gradient issues.

\begin{table}[h]
\centering
\caption{Training Dynamics Metrics}
\label{tab:training-metrics}
\begin{tabular}{lccc}
\toprule
\textbf{Metric} & \textbf{Hybrid} & \textbf{Classical} & \textbf{Interpretation} \\
\midrule
Training Stability Index (TSI) & 0.0025 & 0.0144 & Hybrid is more stable near convergence \\
Training Efficiency Index (TEI) & $\times$ & 0.0000 & Hybrid never reached 90\% accuracy \\
Quantum Gradient Norm (QGN) & 0.458 & --- & Moderate gradient magnitude in last epoch \\
Barren Plateau Indicator (BPI) & 0.0175 & --- & Small but non-vanishing gradients \\
Relative Stability (RQLSI) & 0.1772 & --- & Hybrid model shows lower variance \\
Relative Efficiency (r-QTEI) & $\infty$ & --- & Classical model is significantly faster to train \\
\bottomrule
\end{tabular}
\end{table}

The hybrid model exhibits lower validation loss variability in late training ($\text{TSI} = 0.0025$), indicating consistent behavior, whereas the classical model converges quickly but shows slightly more fluctuation ($\text{TSI} = 0.0144$)



\section{Outlook}
QMetric provides a structured approach for diagnosing hybrid quantum-classical models beyond conventional performance metrics. It highlights key aspects such as training behavior, encoding robustness, and circuit design quality. Future developments will include migration to Qiskit’s Estimator v2\footnote{https://quantum.cloud.ibm.com/docs/en/api/qiskit-ibm-runtime/estimator-v2}, support for additional platforms like PennyLane, and expanded metric coverage for multi-class tasks and generative models.

\section{Availability}
All source code, examples, metric definitions, and plotting utilities are available at
\medskip
\noindent \texttt{\href{https://gitlab.com/illesova.silvie.scholar/qmetric}{https://gitlab.com/illesova.silvie.scholar/qmetric}}

\noindent The repository includes a Conda environment file to reproduce the case study.

\section{Acknowledgments}
MB was supported by Italian Government (Ministero dell’Università e della Ricerca, PRIN 2022 PNRR)---cod.
P2022SELA7: ``RECHARGE: monitoRing, tEsting, and CHaracterization of performAnce Regressions''---D.D. n. 1205 del 28/7/2023.  TR gratefully acknowledges the funding support by program "Excellence initiative—research university" for the AGH University in Krakow as well as the ARTIQ project: UMO-2021/01/2/ST6/00004 and ARTIQ/0004/2021.
\section*{Declaration on Generative AI}
During the preparation of this work, the authors used OpenAI ChatGPT (GPT-4) in order to: assist with text formatting, clarify structure, and edit grammar. After using these tools, the author reviewed and edited the content as needed and takes full responsibility for the publication’s content.

\bibliography{sample-ceur}


\end{document}